\documentclass[a4paper]{article}
\usepackage{verbatim}
\usepackage[utf8]{inputenc}
\usepackage{hyperref}
\usepackage{ifthen}
\usepackage{graphicx}
\usepackage{isabelle,isabellesym}\isabellestyle{literal}

\isadroptag{theory}
\isadroptag{ML}

\renewcommand{\isastyletxt}{\isastyletext}

\newcommand{\isasymSORRY}{$\langle\mathit{proof}\rangle$}
\renewcommand{\isacommand}[1]
{\ifthenelse{\equal{sorry}{#1}}{\;\isasymSORRY}{\isakeyword{#1}}}

\newcommand{\appref}[1]{appendix~\ref{#1}}

\newcommand{\para}[1]{\medskip\noindent\textbf{#1}}

\begin{document}

\title{PIDE as front-end technology for Coq}
\author{Makarius Wenzel \thanks{Current research supported by Project
    Paral-ITP (ANR-11-INSE-001).} \\
  \small Univ. Paris-Sud, Laboratoire LRI, UMR8623, Orsay, F-91405, France \\
  \small CNRS, Orsay, F-91405, France}
\maketitle

\begin{abstract}
  Isabelle/PIDE is the current Prover IDE technology for Isabelle.  It
  has been developed in ML and Scala in the past 4--5 years for this
  particular proof assistant, but with an open mind towards other
  systems.  PIDE is based on an asynchronous document model, where the
  prover receives edits continuously and updates its internal state
  accordingly.  The interpretation of edits and the policies for proof
  document processing are determined by the prover.  The editor
  front-end merely takes care of visual rendering of formal document
  content.

  Here we report on an experiment to connect Coq to the PIDE
  infrastructure of Isabelle.  This requires to re-implement the core
  PIDE protocol layer of Isabelle/ML in OCaml.  The payload for
  semantic processing of proof document content is restricted to
  lexical analysis in the sense of existing CoqIde functionality.
  This is sufficient as proof-of-concept for PIDE connectivity.
  Actual proof processing is then a matter of improving Coq towards
  timeless and stateless proof processing, independently of PIDE
  technicalities.  The implementation worked out smoothly and required
  minimal changes to the refined PIDE architecture of Isabelle2013.

  This experiment substantiates PIDE as general approach to prover
  interaction.  It illustrates how other provers of the greater ITP
  family can participate by following similar reforms of the classic
  TTY loop as was done for Isabelle in the past few years.
\end{abstract}

\begin{isabellebody}%
\def\isabellecontext{Paper}%
\isadelimtheory
\endisadelimtheory
\isatagtheory
\isacommand{theory}\isamarkupfalse%
\ Paper\isanewline
\isakeyword{imports}\ Pure\isanewline
\isakeyword{begin}%
\endisatagtheory
{\isafoldtheory}%
\isadelimtheory
\endisadelimtheory
\isamarkupsection{Motivation%
}
\isamarkuptrue%
\begin{isamarkuptext}%
Is interactive theorem proving inherently tied to the
  command-line?  Is the Proof General wrapper for that command-line
  the optimum of what can be achieved?  Can we ever go beyond it
  conceptually and technologically?

  The PIDE (Prover IDE) approach challenges the predominance of Proof
  General \cite{Aspinall:TACAS:2000} and its many clones like CoqIde,
  Matita \cite{Asperti-et-al:2007}, Proofweb \cite{Kaliszyk:2006}.
  PIDE is centered around general principles of document-oriented
  asynchronous interaction, possibly with parallel processing on the
  prover side \cite{Wenzel:2010,Wenzel:2011:CICM,Wenzel:2012:UITP}.
  It has required several years to reach the first stable release of
  the Isabelle/jEdit Prover IDE in October 2011
  \cite{Wenzel:2012:CICM}.  Isabelle2013 (February 2013) includes the
  third stable release of Isabelle/jEdit, and is already the default
  instead of Proof General (even just for the pragmatic reason that it
  works more likely out-of-the-box).

  Can other provers join this movement?  The present paper reports on
  an experiment called \emph{CoqPIDE} that exchanges the back-end of
  Isabelle/jEdit to use Coq instead of Isabelle (see
  \url{https://bitbucket.org/makarius/coq-pide/src/443d088a72e6/README.PIDE?at=v8.4}
  from January 2013).

  \smallskip The common language of the PIDE prover integration is
  Scala \cite{Scala:2004}, but the prover needs to implement certain
  operations to support document-oriented interaction natively in its
  own language, which is OCaml for Coq.

  Scala provides immediate access to existing Java IDE frameworks.
  Our standard application uses jEdit, but Eclipse, Netbeans, IntelliJ
  IDEA are in principle possible as well.  The JVM is also a good
  platform for advanced web services.

  In the past, OCaml was quite capable to integrate mainstream C
  libraries, such as GTK for GUI components, although that is now
  outdated.  OCaml/GTK was used to implement CoqIde within Coq itself,
  but GTK does not yet provide a full-scaled IDE, not even an able
  text editor.

  The Java platform in general, and jEdit in particular, are not free
  from technical problems, but Isabelle/jEdit shows how these raw
  industrial materials can be adopted for our provers.  Scala is
  particularly helpful to make the JVM platform accessible to the
  higher-order functional culture of proof assistants.

  \smallskip The main results of the CoqPIDE experiment are as
  follows:
  \begin{description}

  \item[Universality.] The requirements for the prover to implement
  the PIDE document model are easily met by porting existing SML
  implementations to OCaml. Note that PIDE defines only rather general
  principles of document editing, leaving most of the details to the
  prover.

  \item[Clarity.] The minimal PIDE protocol implementation for
  OCaml/Coq helps to explain how PIDE actually works, without the
  additional layers of sophistication and performance tuning that have
  accumulated in Isabelle already.

  \item[Frugality.] A meaningful application of PIDE to a different
  prover merely requires 40\,kB of sources in OCaml (26\,kB) and Scala
  (14\,kB).  Approx.\ 50\% is for the implementation of PIDE datatypes
  and protocol operations, the other 50\% Coq-specific ``payload''.
  For more serious semantic processing on the prover side, the payload
  will grow beyond this initial PIDE configuration.

  \end{description}%
\end{isamarkuptext}%
\isamarkuptrue%
\isamarkupsection{PIDE Document Operations%
}
\isamarkuptrue%
\begin{isamarkuptext}%
The PIDE document model maintains sources (produced by the
  editor) and resulting formal content (produced by the prover).  Its
  programming interface consists of statically-typed Scala operations.
  \isa{Document{\isachardot}update{\isacharparenleft}old{\isacharunderscore}version{\isacharcomma}\ new{\isacharunderscore}version{\isacharcomma}\ edits{\isacharparenright}} applies
  source edits to turn one version non-destructively into another.
  \isa{Document{\isachardot}remove{\isacharunderscore}versions{\isacharparenleft}versions{\isacharparenright}} indicates obsolete
  versions to allow garbage collection eventually.
  
  Document update is declarative: it specifies where to insert or
  remove parts of the source text, but its operational consequences
  are determined by the prover.  Each document version is associated
  with an \emph{execution} in ML to work out these formal details, and
  report results back to the Scala side.  PIDE provides operational
  hints to improve performance, like \isa{Document{\isachardot}discontinue{\isacharunderscore}execution{\isacharparenleft}{\isacharparenright}} and \isa{Document{\isachardot}cancel{\isacharunderscore}execution{\isacharparenleft}{\isacharparenright}} in certain situations of its editing
  pipeline.  Cancellation should cause some physical interrupt within
  the prover, which is important for long-running proof checking, but
  the CoqPIDE ignores this for now.

  Note that the PIDE model is inherently asynchronous: the front-end
  never waits for the back-end.  Uninterruptible execution could mean
  (infinitely) long delay of the update of formal annotations seen in
  the editor buffer, but the Prover IDE does not block, nor lock text
  in the manner of Proof General.%
\end{isamarkuptext}%
\isamarkuptrue%
\isamarkupsection{PIDE Protocol Implementation (OCaml)%
}
\isamarkuptrue%
\begin{isamarkuptext}%
The PIDE protocol merely propagates tree-structured datatypes
  between Scala and ML, but various details need to be observed to
  make it robust, efficient, and portable.  It is a bit like plumbing
  different kinds of metal: a leaden JVM with an aluminium ML system,
  using a copper pipe.  The PIDE protocol stack has evolved over
  several years to the ``proven technology'' in Isabelle2013.  For
  CoqPIDE, we re-implemented the ML side in OCaml, which turned out a
  simple programming exercise of a few days (including to learn some
  OCaml in the first place).  The main protocol layers are as follows.

  \para{Bidirectional byte-channel.} PIDE demands a bidirectional
  communication channel based on clean byte-streams, with
  block-buffering and high throughput.  On Unix this can be
  implemented by a pair of fifos (named pipes), which are opened like
  a regular file on each side, in the correct order for rendevouz.
  PIDE on Windows uses TCP sockets instead, but they turn out slightly
  less efficient and less robust on some ML versions.  For CoqPIDE we
  use fifos and thus restrict it to Unix for now, although OCaml
  sockets probably work as well.

  Content sent over the byte-channel is partitioned into \emph{chunks}
  with explicit length indication (encoded via ASCII digits followed
  by newline).  This depends on the assumption that the channel is
  \emph{private} to the protocol handler.

  Note that public \texttt{stdout} cannot be used, because it is
  subject to spurious output by parts of the ML process (runtime
  system, libraries) beyond our control.  Classic Proof General
  \cite{Aspinall:TACAS:2000} avoids this problem by using
  human-readable control commands and asking the user for manual
  repairs when the protocol looses synchronization.  This no longer
  worked for the PGIP/XML protocol \cite{Aspinall-et-al:2007}, so it
  was suffering from breakdown caused by unexpected diagnostic
  messages.

  \para{Text encoding and character positions.} Text on the JVM
  consists of 16-bit characters, but requires one or two such
  characters to represent a single Unicode 6 codepoint according to
  UTF-16.  ML usually prefers some extension of ASCII, formerly
  ISO-latin, now UTF-8 where multi-character encodings are
  commonplace.  In any case, logical text addressing needs to agree on
  both sides to attach error messages or other formal markup precisely
  to source positions.

  PIDE standardizes towards UTF-8 on the prover side and recodes text
  to UTF-16 for Scala/JVM.  Physical text addressing works via byte
  offsets in ML, and character offsets on the JVM.  Logical text
  positions are either translated explicitly by functions provided by
  the prover, or represented in a way that is invariant wrt.\ the
  encoding (like Isabelle symbols \cite[\S2.1]{Wenzel:2011:CICM}).
  For CoqPIDE we have re-used \verb,byte_offset_to_char_offset, from
  CoqIde, which works for the \emph{Basic Multilingual Plane} where
  UTF-16 requires only one 16-bit character.

  \para{YXML transfer syntax of untyped trees.} PIDE uses untyped XML
  trees for document markup (and arbitrary ML/Scala values), but
  ignores the complications of official XML syntax and various
  attempts at XML type-systems.  Instead, the markup tree structure
  over the text is represented by two special control characters that
  are outside the text range of XML 1.0 and what provers normally use.
  In contrast to official XML syntax, this avoids quoting of the text
  and allows cumulative markup of text that might have been marked
  already.

  Our XML transfer syntax is called YMXL (pronounced as ``Why XML?''),
  see also \cite[\S2.3]{Wenzel:2011:CICM}.  Efficient and robust YXML
  parsing is easily implemented in any programming language.  What is
  also notable about YXML is that it is orthogonal to UTF-8 text
  encoding: the operations to decode text and to recover tree
  structure can be \emph{commuted} by the PIDE infrastructure as
  required.

  The OCaml version of YXML is a literal translation of the SML
  version from Isabelle, see also
  \url{https://bitbucket.org/makarius/yxml} and \appref{app:yxml}.
  Note that CoqIde uses some \emph{XML Light} implementation instead,
  which suffers from typical problems with boundary cases of standard
  XML (e.g.\ incorrect treatment of white-space).

  \para{XML/ML data representation.}  The algebraic datatypes that
  PIDE transfers between Scala and ML may consist of base types like
  \isa{bool}, \isa{int}, \isa{string} (for text in the
  above sense), product types (tuples or records), variant types
  (disjoint sums), and recursion over the same.  Imitating the
  canonical memory layout of ML values in untyped memory, we provide
  ML functions and combinators for each of these type constructions
  wrt.\ raw XML trees:

{\footnotesize
\begin{verbatim}
type 'a Encode.t = 'a -> XML.tree list
Encode.string: string Encode.t
Encode.pair: 'a Encode.t -> 'b Encode.t -> ('a * 'b) Encode.t
Encode.list: 'a Encode.t -> 'a list Encode.t
\end{verbatim}
}

  \noindent Etc., also with symmetric versions for \verb,Decode,.  The
  modules \verb,XML.Encode, and \verb,XML.Decode, are available on the
  Scala side as well.  The implementation is mostly trivial,
  consisting of a few lines for each combinator.  Note that it is
  important to work recursively with \verb,XML.tree list, and cope
  with its ``mixed content'' of alternating \verb,XML.Elem, and
  \verb,XML.Text, nodes.  See also
  \url{https://bitbucket.org/makarius/yxml} and \appref{app:xml}.

  Each PIDE protocol function is wrapped into a combinator expression
  over \verb,XML.Encode, and \verb,XML.Decode, that is isomorphic to
  the corresponding datatype definitions of its arguments.  This
  slight redundancy is isolated in a single place in Scala and ML,
  respectively.  Public interfaces on each side are statically
  typed.%
\end{isamarkuptext}%
\isamarkuptrue%
\isamarkupsection{Coq-specific PIDE Modules (OCaml and Scala)%
}
\isamarkuptrue%
\begin{isamarkuptext}%
After studying the Coq sources for a few days, to see what is
  already there to serve our purpose, we have chosen lexical syntax
  processing of CoqIde: it is used for syntax highlighting in its GTK
  text widget.

  \begin{figure}[htb]
  \centering
  \includegraphics[width=0.85\textwidth]{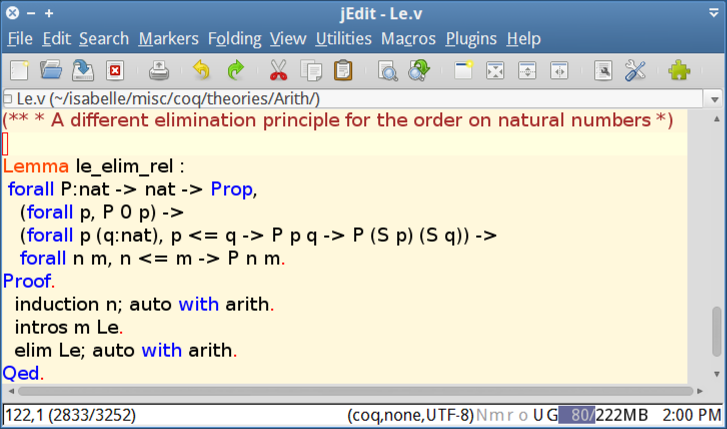}
  \end{figure}
  
  \noindent The result looks like in the given screenshot, but that is
  already CoqPIDE imitating the look-and-feel of CoqIde! This minimal
  PIDE application has required the following addtional modules.

  \para{Coq document structure.} CoqPIDE implements the main PIDE
  document operations using a very simple document model in OCaml: an
  association list of named nodes (\verb,.v, files), each consisting
  on a list of entries (like command spans in Proof~General).  The
  Scala-side of CoqPIDE does not even exploit the full structure yet:
  it merely turns each file into one monolithic command span.  This is
  sufficient to work efficiently with files of 10--100\,kB size.
  Editing larger sources requires to use more sub-structure as is done
  routinely in Isabelle/PIDE.

  \para{Coq markup and rendering.}  PIDE allows the prover to annotate
  source text by arbitrary formal content, which is represented by
  untyped and uninterpreted XML trees.  The prover may define its own
  vocabulary of \emph{markup} elements, together with an
  interpretation that is called \emph{rendering}.  The rendering
  module provides functions to turn XML trees that are associated with
  given text ranges into GUI elements: colors, boxes, squiggles,
  icons, tooltips, hyperlinks etc.

  CoqPIDE defines markup elements for the main lexical categories of
  Coq, with rendering that uses the original CoqIde colors.  There are
  two tiny additions: the dot as proof-script terminator is painted
  red, and quoted strings are rendered with transparency (alpha
  channel) which is now common-place in Isabelle/PIDE --- it helps to
  combine multiple layers of prover markup systematically.

  \para{Coq theory syntax.} PIDE allows the prover to define aspects
  of its syntax directly on the editor side, e.g.\ to tell which files
  are special (\verb,.v, for Coq) and how files are loaded for the
  prover (sources are managed by the front-end).  CoqPIDE only
  provides the bare minimum to make the system work.  Extra efforts
  would be required to imitate Isabelle/PIDE, which allows to augment
  theory syntax while editing, and resolve file dependencies
  automatically.  The latter is important for applications consisting
  of several modules that are edited simultaneously.  Note that
  neither CoqIde nor Proof General allow multiple ``active'' buffers.

  \smallskip Of course we could have augmented jEdit directly by a
  small Coq lexer in Scala to get plain syntax highlighting.  Since
  CoqPIDE follows the architecture of PIDE, with its asynchronous
  exchange of edits that are interpreted by the prover, it may serve
  as a starting point for actual proof processing.  Some of the
  required reforms to the Coq toplevel have been explored already
  \cite{Tassi-Barras:2012}.  Eventually we shall see these movements
  converge to more serious PIDE/Coq integration.

  Isabelle shares the roots of TTY-based command-line interaction with
  Coq and other members of the LCF family, but managed to reform
  itself towards full-scale Prover IDE support within a few years.  It
  should be feasible to transfer more of what has been achieved here
  to other provers, unless the command-line really turns out as part
  of the very essence of interactive theorem proving.%
\end{isamarkuptext}%
\isamarkuptrue%
\isadelimtheory
\endisadelimtheory
\isatagtheory
\isacommand{end}\isamarkupfalse%
\endisatagtheory
{\isafoldtheory}%
\isadelimtheory
\endisadelimtheory
\isanewline
\end{isabellebody}%
%%% Local Variables:
%%% mode: latex
%%% TeX-master: "root"
%%% End:

\appendix

\section{OCaml sources of the YXML library}

\subsection{YXML transfer syntax \label{app:yxml}}

{\small
\begin{verbatim}
(*
Efficient text representation of XML trees using extra characters X
and Y -- no escaping, may nest marked text verbatim.  Suitable for
direct inlining into plain text.

Markup <elem att="val" ...>...body...</elem> is encoded as:

  X Y name Y att=val ... X
  ...
  body
  ...
  X Y X
*)

module type YXML =
sig
  val char_X: char
  val char_Y: char
  val no_output: string * string
  val output_markup: string * XML.attributes -> string * string
  val string_of_body: XML.body -> string
  val string_of: XML.tree -> string
  val parse_body: string -> XML.body
  val parse: string -> XML.tree
end

module YXML: YXML =
struct

(* markers *)

let char_X = '\005'
let char_Y = '\006'

let str_X = String.make 1 char_X
let str_Y = String.make 1 char_Y

let str_XY = str_X ^ str_Y
let str_XYX = str_XY ^ str_X

let detect s = String.contains s char_X || String.contains s char_Y


(* ML basics *)

let (|>) x f = f x
let (@>) f g x = g (f x)

let rec fold f list y =
  match list with
    [] -> y
  | x :: xs -> fold f xs (f x y)


(* output *)

let implode = String.concat ""
let content xs = implode (List.rev xs)
let add x xs = if x = "" then xs else x :: xs

let no_output = ("", "")

let output_markup (name, atts) =
  if name = "" then no_output
  else
    (str_XY ^ name ^
      implode (List.map (fun (a, x) -> str_Y ^ a ^ "=" ^ x) atts) ^ str_X,
     str_XYX)

let string_of_body body =
  let attrib (a, x) = add str_Y @> add a @> add "=" @> add x in
  let rec tree = function
    | XML.Elem ((name, atts), ts) ->
        add str_XY @> add name @> fold attrib atts @> add str_X @>
        trees ts @>
        add str_XYX
    | XML.Text s -> add s
  and trees ts = fold tree ts
  in content (trees body [])

let string_of tree = string_of_body [tree]



(** parsing *)

(* split *)

let split fields sep str =
  let cons i n result =
    if i = 0 && n = String.length str && n > 0 then str :: result
    else if n > 0 then String.sub str i n :: result
    else if fields then "" :: result
    else result
  in
  let rec explode i result =
    let j = try String.index_from str i sep with Not_found -> -1 in
      if j >= 0 then explode (j + 1) (cons i (j - i) result)
      else List.rev (cons i (String.length str - i) result)
  in explode 0 []


(* parse *)

let err msg = raise (Failure ("Malformed YXML: " ^ msg))
let err_attribute () = err "bad attribute"
let err_element () = err "bad element"
let err_unbalanced name =
  if name = "" then err "unbalanced element"
  else err ("unbalanced element \"" ^ name ^ "\"")

let parse_attrib s =
  try
    let i = String.index s '=' in
    let _ = if i = 0 then err_attribute () in
    let j = i + 1 in
      (String.sub s 0 i, String.sub s j (String.length s - j))
  with Not_found -> err_attribute ()

let parse_body source =

  (* stack operations *)

  let add x ((elem, body) :: pending) = (elem, x :: body) :: pending
  in

  let push name atts pending =
    if name = "" then err_element ()
    else ((name, atts), []) :: pending
  in

  let pop (((name, _) as markup, body) :: pending) =
    if name = "" then err_unbalanced ""
    else add (XML.Elem (markup, List.rev body)) pending
  in

  (* parse chunks *)

  let chunks = split false char_X source |> List.map (split true char_Y) in

  let parse_chunk = function
    | [""; ""] -> pop
    | ("" :: name :: atts) -> push name (List.map parse_attrib atts)
    | txts -> fold (fun s -> add (XML.Text s)) txts
  in
  match fold parse_chunk chunks [(("", []), [])] with
  | [(("", _), result)] -> List.rev result
  | ((name, _), _) :: _ -> err_unbalanced name

let parse source =
  match parse_body source with
  | [result] -> result
  | [] -> XML.Text ""
  | _ -> err "multiple results"

end

\end{verbatim}
}

\subsection{Untyped XML trees and typed representation of ML
  values \label{app:xml}}

{\small
\begin{verbatim}
module type XML_Data_Ops =
sig
  type 'a a
  type 'a t
  type 'a v
  val int_atom: int a
  val bool_atom: bool a
  val unit_atom: unit a
  val properties: (string * string) list t
  val string: string t
  val int: int t
  val bool: bool t
  val unit: unit t
  val pair: 'a t -> 'b t -> ('a * 'b) t
  val triple: 'a t -> 'b t -> 'c t -> ('a * 'b * 'c) t
  val list: 'a t -> 'a list t
  val option: 'a t -> 'a option t
  val variant: 'a v list -> 'a t
end

module type XML =
sig
  type attributes = (string * string) list
  type tree = Elem of ((string * attributes) * tree list) | Text of string
  type body = tree list
  exception XML_Atom of string
  exception XML_Body of tree list
  module Encode: XML_Data_Ops with
    type 'a a = 'a -> string and
    type 'a t = 'a -> body and
    type 'a v = 'a -> string list * body
  module Decode: XML_Data_Ops with
    type 'a a = string -> 'a and
    type 'a t = body -> 'a and
    type 'a v = string list * body -> 'a
end

module XML: XML =
struct

type attributes = (string * string) list
type tree = Elem of ((string * attributes) * tree list) | Text of string
type body = tree list


let map_index f =
  let rec mapp i = function
    | [] -> []
    | x :: xs -> f (i, x) :: mapp (i + 1) xs
  in mapp 0

exception XML_Atom of string
exception XML_Body of tree list


module Encode =
struct

type 'a a = 'a -> string
type 'a t = 'a -> body
type 'a v = 'a -> string list * body


(* atomic values *)

let int_atom = string_of_int

let bool_atom = function false -> "0" | true -> "1"

let unit_atom () = ""


(* structural nodes *)

let node ts = Elem ((":", []), ts)

let vector = map_index (fun (i, x) -> (int_atom i, x))

let tagged (tag, (xs, ts)) = Elem ((int_atom tag, vector xs), ts)


(* representation of standard types *)

let properties props = [Elem ((":", props), [])]

let string = function "" -> [] | s -> [Text s]

let int i = string (int_atom i)

let bool b = string (bool_atom b)

let unit () = string (unit_atom ())

let pair f g (x, y) = [node (f x); node (g y)]

let triple f g h (x, y, z) = [node (f x); node (g y); node (h z)]

let list f xs = List.map (fun x -> node (f x)) xs

let option f = function None -> [] | Some x -> [node (f x)]

let variant fns x =
  let rec get_index i = function
    | [] -> raise (Failure "XML.Encode.variant")
    | f :: fs -> try (i, f x) with Match_failure _ -> get_index (i + 1) fs
  in [tagged (get_index 0 fns)]

end


module Decode =
struct

type 'a a = string -> 'a
type 'a t = body -> 'a
type 'a v = string list * body -> 'a


(* atomic values *)

let int_atom s =
  try int_of_string s
    with Invalid_argument _ -> raise (XML_Atom s)

let bool_atom = function
  | "0" -> false
  | "1" -> true
  | s -> raise (XML_Atom s)

let unit_atom s =
  if s = "" then () else raise (XML_Atom s)


(* structural nodes *)

let node = function
  | Elem ((":", []), ts) -> ts
  | t -> raise (XML_Body [t])

let vector =
  map_index (function (i, (a, x)) ->
    if int_atom a = i then x else raise (XML_Atom a))

let tagged = function
  | Elem ((name, atts), ts) -> (int_atom name, (vector atts, ts))
  | t -> raise (XML_Body [t])


(* representation of standard types *)

let properties = function
  | [Elem ((":", props), [])] -> props
  | ts -> raise (XML_Body ts)

let string = function
  | [] -> ""
  | [Text s] -> s
  | ts -> raise (XML_Body ts)

let int ts = int_atom (string ts)

let bool ts = bool_atom (string ts)

let unit ts = unit_atom (string ts)

let pair f g = function
  | [t1; t2] -> (f (node t1), g (node t2))
  | ts -> raise (XML_Body ts)

let triple f g h = function
  | [t1; t2; t3] -> (f (node t1), g (node t2), h (node t3))
  | ts -> raise (XML_Body ts)

let list f = List.map (fun t -> f (node t))

let option f = function
  | [] -> None
  | [t] -> Some (f (node t))
  | ts -> raise (XML_Body ts)

let variant fs = function
  | [t] ->
      let (tag, (xs, ts)) = tagged t in
      let f = try List.nth fs tag
        with Invalid_argument _ -> raise (XML_Body [t]) in
      f (xs, ts)
  | ts -> raise (XML_Body ts)

end

end

\end{verbatim}
}

\bibliographystyle{abbrv}
\bibliography{root}

\begin{thebibliography}{10}

\bibitem{Asperti-et-al:2007}
A.~Asperti, C.~Sacerdoti~Coen, E.~Tassi, and S.~Zacchiroli.
\newblock User interaction with the {Matita} proof assistant.
\newblock {\em Journal of Automated Reasoning}, 39(2), 2007.

\bibitem{Aspinall:TACAS:2000}
D.~Aspinall.
\newblock {Proof General}: A generic tool for proof development.
\newblock In S.~Graf and M.~Schwartzbach, editors, {\em European Joint
  Conferences on Theory and Practice of Software (ETAPS)}, volume 1785 of {\em
  LNCS}. Springer, 2000.

\bibitem{Aspinall-et-al:2007}
D.~Aspinall, C.~L\"uth, and D.~Winterstein.
\newblock A framework for interactive proof.
\newblock In M.~Kauers, M.~Kerber, R.~Miner, and W.~Windsteiger, editors, {\em
  Towards Mechanized Mathematical Assistants (CALCULEMUS and MKM 2007)}, volume
  4573 of {\em LNAI}. Springer, 2007.

\bibitem{Kaliszyk:2006}
C.~Kaliszyk.
\newblock Web interfaces for proof assistants.
\newblock In S.~Autexier and C.~Benzm\"uller, editors, {\em User Interfaces for
  Theorem Provers (UITP 2006)}, volume 174(2) of {\em ENTCS}. Elsevier, 2007.

\bibitem{Scala:2004}
M.~Odersky et~al.
\newblock An overview of the {Scala} programming language.
\newblock Technical Report IC/2004/64, EPF Lausanne, 2004.

\bibitem{Tassi-Barras:2012}
E.~Tassi and B.~Barras.
\newblock Designing a state transaction machine for {Coq}.
\newblock In {\em The Coq Workshop 2012 (co-located with ITP 2012)}, 2012.

\bibitem{Wenzel:2010}
M.~Wenzel.
\newblock Asynchronous proof processing with {Isabelle/Scala} and
  {Isabelle/jEdit}.
\newblock In C.~Sacerdoti~Coen and D.~Aspinall, editors, {\em User Interfaces
  for Theorem Provers (UITP 2010)}, ENTCS, July 2010.
\newblock FLOC 2010 Satellite Workshop.

\bibitem{Wenzel:2011:CICM}
M.~Wenzel.
\newblock Isabelle as document-oriented proof assistant.
\newblock In J.~H. Davenport, W.~M. Farmer, F.~Rabe, and J.~Urban, editors,
  {\em Conference on Intelligent Computer Mathematics / Mathematical Knowledge
  Management (CICM/MKM 2011)}, volume 6824 of {\em LNAI}. Springer, 2011.

\bibitem{Wenzel:2012:CICM}
M.~Wenzel.
\newblock {Isabelle/jEdit} --- a {Prover IDE} within the {PIDE} framework.
\newblock In J.~Jeuring et~al., editors, {\em Conference on Intelligent
  Computer Mathematics (CICM 2012)}, volume 7362 of {\em LNAI}. Springer, 2012.

\bibitem{Wenzel:2012:UITP}
M.~Wenzel.
\newblock {READ-EVAL-PRINT} in parallel and asynchronous proof-checking.
\newblock In {\em User Interfaces for Theorem Provers (UITP 2012)}, EPTCS,
  2013.

\end{thebibliography}

\end{document}